\documentclass[10pt,times,a4paper]{iopart}
\usepackage{epsfig}
\usepackage{graphics}
\usepackage{graphicx}
\usepackage{bm}
\usepackage{iopams}
\usepackage{color}
\usepackage{bm}

\begin{document}

\newcommand{\rhat}{\hat{r}}
\newcommand{\iotahat}{\hat{\iota}}
\newcommand{\phihat}{\hat{\phi}}
\newcommand{\hc}{\textsf{h}}
\newcommand{\be}{\begin{equation}}
\newcommand{\ee}{\end{equation}}
\newcommand{\ber}{\begin{eqnarray}}
\newcommand{\eer}{\end{eqnarray}}
\newcommand{\fmerg}{f_{\rm merg}}
\newcommand{\fcut}{f_{\rm cut}}
\newcommand{\fring}{f_{\rm ring}}
\newcommand{\cA}{\mathcal{A}}
\newcommand{\ie}{i.e.}

\newcommand{\df}{{\mathrm{d}f}}
\newcommand{\dt}{{\mathrm{d}t}}
\newcommand{\pj}{\partial_j}
\newcommand{\pk}{\partial_k}
\newcommand{\psifl}{\Psi(f; {\bm \lambda})}
\newcommand{\romani}{\mathrm{i}}

\title[Gravitational-wave data analysis using binary black-hole waveforms]
{Gravitational-wave data analysis using binary black-hole waveforms}

\author{P.~Ajith}
\address{Max-Planck-Institut f\"ur Gravitationsphysik 
(Albert-Einstein-Institut) and Leibniz Universit\"at Hannover, 
Callinstr.~38, 30167~Hannover, Germany}

\begin{abstract}
While the \emph{inspiral} and \emph{ring-down} stages of the binary 
black-hole coalescence are well-modelled by analytical approximation methods in general relativity, 
the recent progress in numerical relativity has enabled us to compute accurate waveforms from the
\emph{merger} stage also. This has an important impact on the search for gravitational waves from binary 
black holes. `Complete' binary black-hole waveforms can now be produced by matching post-Newtonian 
waveforms with those computed by numerical relativity, which can be parametrised to produce analytical 
waveform templates. The `complete' waveforms can also be used to estimate the efficiency of different 
search methods aiming to detect signals from black-hole coalescences. This paper summarises some 
recent efforts in this direction.
\end{abstract}

\section{Introduction}

Coalescing black-hole binaries are among the most promising sources of gravitational waves
for ground-based detectors like LIGO and Virgo, and the planned space-borne detector LISA. The 
evolution of binary black holes is conventionally split into three stages: \emph{inspiral, merger} 
and \emph{ring down}. In the inspiral stage, the two compact objects, driven by radiation reaction, 
move in quasi-circular orbits. Eventually approaching the ultra-relativistic regime, the two bodies 
merge to form a single excited Kerr black hole. In the ring-down stage, the excited black hole loses 
its energy by gravitational-wave emission and settles into a Kerr black hole. Gravitational waveforms from  the 
inspiral and ring-down stages can be accurately computed by approximation/perturbation techniques in 
general relativity~\cite{BDEI04,ABIQ04,TeukolskyandPress:1974}. The recent progress in numerical relativity
\cite{Pretorius:2005gq,Campanelli:2005dd,Baker05a} has enabled us to model also the non-perturbative 
merger phase of the coalescence of binary black holes~\cite{Baker2006b,Gonzalez06tr,Campanelli2006b,
Campanelli2006c,Herrmann:2007ac,Koppitz-etal-2007aa,Gonzalez:2007hi,Campanelli2007,Campanelli:2006fy,
Pollney:2007ss,Rezzolla-etal-2007}.

While the current gravitational-wave searches look for each stage of the binary black-hole coalescence separately
(see, for example,~\cite{Abbott:2007xi,Goggin:2006bs}), combining the results from analytical and 
numerical relativity enables us to \emph{coherently} search for all the three stages using a single 
template family. This coherent search is significantly more sensitive than the current searches over 
certain mass ranges (see Section~\ref{sec:searchCompare}). This search has added advantages: including 
all the three stages adds more `structure' to the template waveform, resulting in a potential reduction of false 
alarms. The additional structure and the improved signal-to-noise ratio also results in an 
improved estimation of the parameters of the binary, which is particularly important for 
LISA data analysis. As LISA data will contain a `cocktail' of many strong binary signals, these will 
have to be subtracted from the data in order to analyse other signals. Improved parameter estimation 
can also have a tremendous impact in cosmology. Since many of the supermassive black-hole mergers 
are likely to have electromagnetic counterparts, it is possible to constrain the values of 
cosmological parameters by combining the gravitational-wave and electromagnetic observations~\cite{Schutz:1986}. 
In particular, using the distance-redshift relation from many binary black-hole `\emph{standard sirens}', 
LISA might be able to put interesting constraints on the equation of state of the dark 
energy~\cite{HolzHughes:2005}. The error bars on this depend on how accurately the 
`red-shifted' mass of the source and the luminosity distance are estimated, and how well the host 
galaxy of the electromagnetic counterpart is identified. The improved parameter estimation might help to tighten 
these constraints. 

Several authors have proposed different ways of computing gravitational-wave templates containing 
all the three stages of the binary black-hole coalescence~\cite{Damour:2007xr,Buonanno:2007pf,
Pan:2007nw,Ajith:2007kx,Ajith:2007qp}. In particular, Ref.~\cite{Ajith:2007kx} proposed a phenomenological 
parametrisation for non-spinning binary black-hole waveforms. These waveforms are explicit functions of the 
physical parameters of the system and exhibited very high overlaps with the `target signals'. The target
signals were constructed by matching numerical-relativity waveforms with post-Newtonian waveforms in 
appropriate matching regions. These waveforms contain all the three stages of the binary black-hole coalescence. 
Hence, they can also be used to estimate the efficiency of different (template-based and other)
search methods used to detect signals from binary black-hole coalescences. 

This paper provides an overview of the application of the results from analytical and numerical
calculations of binary black-hole waveforms into gravitational-wave data analysis. Section~\ref{sec:hybWave}
describes how `complete' binary black-hole coalescence waveforms (so-called hybrid waveforms) from 
non-spinning binaries can be constructed by matching post-Newtonian and numerical-relativity waveforms. 
Here we also investigate the robustness of the matching procedure by studying the mismatch between
hybrid waveforms constructed using different matching regions. 
Section~\ref{sec:phenTempl} introduces an analytical two-parameter family of non-spinning waveforms
having very good overlaps with the hybrid waveforms. Section~\ref{sec:searchCompare} shows how
the hybrid waveforms can be used to estimate the sensitivity of different searches. Here we compare
the \emph{fitting factor} and \emph{faithfulness} of different template-based searches using the hybrid 
waveforms as target signals. Finally, Section~\ref{sec:Summary} provides a summary and future plans.

\section{Constructing `complete' binary black-hole coalescence signals}
\label{sec:hybWave}

\begin{figure}[tb]
\centering
\includegraphics[width=5.5in]{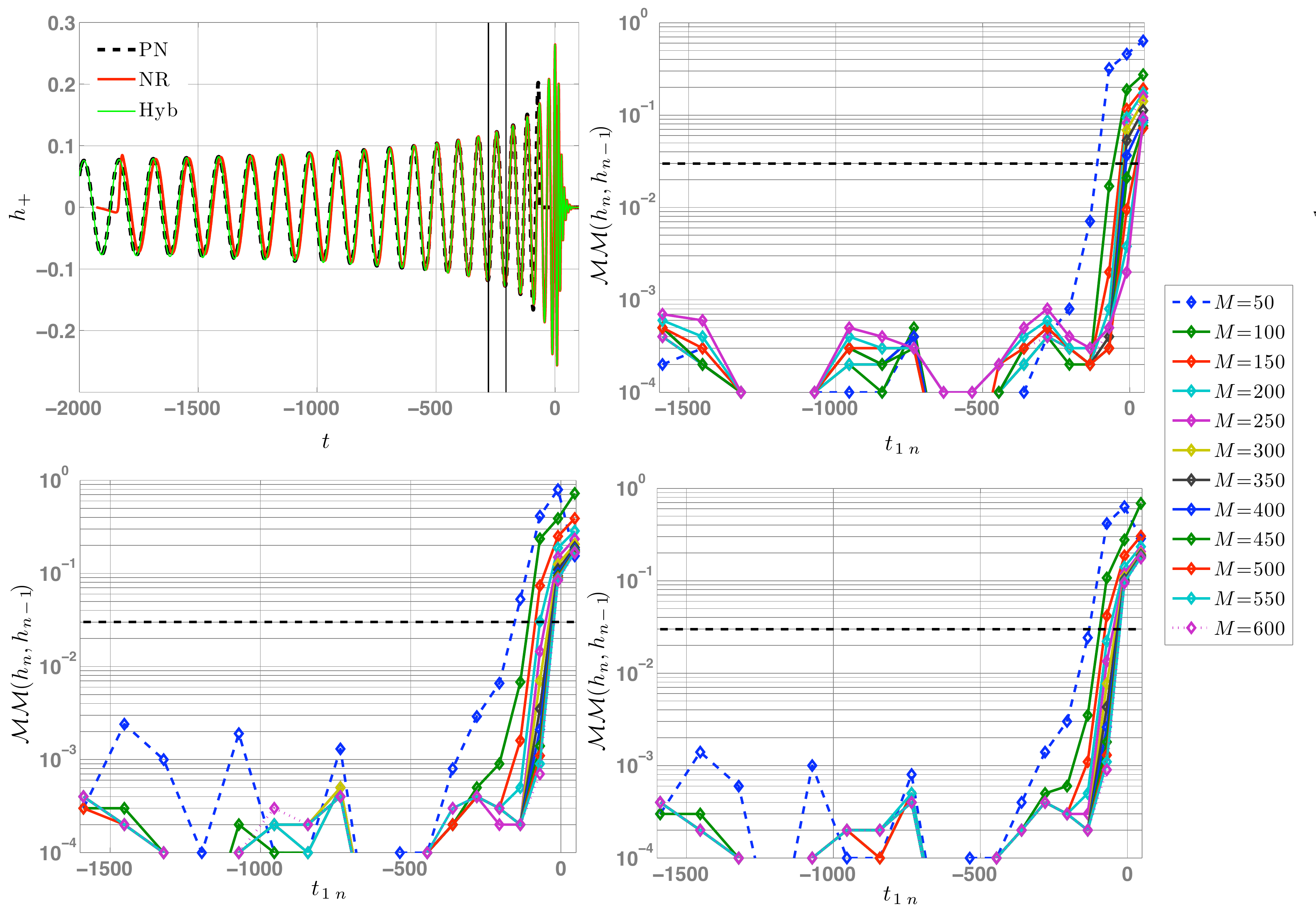}
\caption{Top-left panel shows an example hybrid waveform (green) constructed by matching 
an equal-mass NR waveform (red) computed by the Jena group with a 3.5PN restricted PN 
waveform (black dashed). The two black vertical lines indicate the matching region ($-280 M \leq t \leq 
-206 M$) employed. The rest of the panels show the mismatch between hybrid waveforms
$h_n$ and $h_{n-1}$, where a subscript $n$ means that the hybrid waveform is constructed by 
matching the NR and PN waveforms at the $n\,$th cycle of the NR waveform. The horizontal
axis reports the start time, $t_1$, of the $n\,$th cycle (in units of $M$). The mismatch 
is computed using Initial LIGO (top-right), Advanced LIGO (bottom-right) and Virgo (bottom-left) 
noise spectra. Total mass of the waveforms (in units of $M_\odot$) is shown in the legends.
A mismatch of 3\% is marked with a dashed horizontal line.}
\label{fig:HybWaveOverlaps}
\end{figure}
%
Although numerical relativity (NR) is able to compute gravitational waveforms containing all the 
three stages of the binary black-hole coalescence, the numerical simulations are heavily limited by the 
computational resources. But, the post-Newtonian (PN) formalism is known to work very well in
the early inspiral. Thus, complete binary black-hole coalescence waveforms (hybrid 
waveforms) can be constructed by matching PN and NR waveforms in an appropriate matching region. 
Different authors have studied the consistency of the non-spinning PN waveforms with NR waveforms. 
See Refs.~\cite{Baker:2006ha,Buonanno:2006ui,Berti:2007fi,Pan:2007nw,Ajith:2007qp,
Hannam:2007ik,Boyle:2007ft,Ajith:2007kx,Ajith:2007qp} for some of the recent work.

The time-domain waveform $h_{+,\times}(t,{\bm \mu})$ from a particular system is parametrised 
by a set of `extrinsic parameters' ${\bm \mu} = \{\varphi_0, t_0\}$, where $\varphi_0$ 
is the initial phase and $t_0$ is the start time of the waveform. The PN and NR waveforms are 
matched by minimising the integrated squared difference, $\delta$, between them in
the matching region $t_1 \leq t \leq t_2$:
\ber
\delta &\equiv& \sum_{i=+,\times} \, \int_{t_1}^{t_2}  
\left[h_i^{^{\rm PN}}(t,{\bm \mu})-a\, h_i^{^{\rm NR}}(t,{\bm \mu})\right]^2 \, \dt.
\eer
The minimisation is carried out over the extrinsic parameters ${\bm \mu}$ of the PN waveform 
and an amplitude scaling factor $a$~\footnote{The NR waveforms used for this work contains 
only the $l=2,m=\pm2$ modes, and the PN waveforms are computed in the \emph{restricted}
PN approximation. Since the PN corrections to the amplitude of the PN waveforms are ignored,
this will introduce an error of $\sim8\%$ in the amplitude of the hybrid waveforms, and hence 
in the horizon distances reported in Figure~\ref{fig:HorizonDistance};
but not in the calculation of fitting factor and faithfulness.}, 
while keeping the `intrinsic parameters' (the two component
masses) of both the PN and NR waveforms the same. The hybrid waveforms are then produced by 
combining the `best-matched' PN waveforms with the NR waveforms in the following way:
\be
h_{+,\times}^{{\rm hyb}}(t, {\bm \mu}) \equiv a_0 \, \tau(t) \, h_{+,\times}^{^{\rm NR}}(t, {\bm \mu})
+(1-\tau(t)) \, h_{+,\times}^{^{\rm PN}}(t, {\bm \mu_0})
\label{eq:HybWave}
\ee
where $\bm \mu_0$ and $a_0$ denote the values of $\bm \mu$ and $a$ for
which $\delta$ is minimised, and $\tau$ is a weighting function, defined as
\ber
\tau(t) \equiv \left\{ \begin{array}{ll}
0 & \textrm{if $t < t_1 $}\\ 
\frac{t-t_1}{t_2-t_1}  & \textrm{if $t_1 \leq t < t_2 $}\\ 
1 & \textrm{if $t_2 \leq t$.}
\end{array} \right.
\label{eq:HybWaveWeight}
\eer

It is expected that the early inspiral is better-modelled by PN waveforms, 
as the early-inspiral part of the NR waveforms are prone to larger errors. But the PN 
waveforms become less accurate at the late inspiral. Thus, it is important to choose a matching 
region where both waveforms are accurate. This also means that the hybrid waveforms constructed 
using two very different matching regions can potentially be quite different. If, on the other 
hand, we are able to show that these differences are not very significant for data-analysis 
purpose, this is an indication that our analysis is not heavily dependent on the choice of 
the matching region.

As a test of the robustness of the matching procedure, we compute the \emph{mismatch} between
two hybrid waveforms $h_n$ and $h_{n-1}$, defined as
\be
\mathcal{MM} (h_n,h_{n-1}) \equiv 1 - \mathrm{max}_{t_0} \left[ 4 \, \mathrm{Re} \int_0^\infty 
\frac{h_{n}(f)\, h^*_{n-1}(f) \, e^{\romani 2\pi ft_0}\, \df} {S_h(f)} \right],
\ee 
where $S_h(f)$ is the one-sided power spectral density of the detector noise. The subscript
$n$ on the hybrid waveform $h$ means that the hybrid waveform is constructed by matching PN
and NR waveforms at the $n$th cycle of the NR waveform.
 
The top-left panel of Figure~\ref{fig:HybWaveOverlaps} shows an example set of the PN, NR and the 
hybrid waveforms. The hybrid waveform is constructed by matching an equal-mass NR waveform 
computed by the Jena group, reported in Ref.~\cite{Hannam:2007ik}, with a \emph{restricted} 3.5PN 
TaylorT1~\cite{DIS01} waveform. Other panels in Figure~\ref{fig:HybWaveOverlaps} show the mismatch  
between the hybrid waveforms $h_n$ and $h_{n-1}$, computed using three different noise spectra.
If we take 3\% as the maximum allowed mismatch between hybrid waveforms $h_n$ and $h_{n-1}$, 
this preliminary exercise suggests that any matching region before $t_1 = -150 M$ is robust for 
constructing hybrid waveforms using the equal-mass NR waveforms considered here. This will be 
studied in detail in a forthcoming work~\cite{NRDApaper3}.

\section{Templates for binary black-hole coalescence}
\label{sec:phenTempl}

The hybrid waveforms constructed in the previous section can be parametrised in terms of the
two physical parameters of the binary, thus producing analytical waveform templates. These 
analytical waveforms can be used to construct template banks for matched-filter searches; thus avoiding the 
computational cost of generating hybrid waveforms at each grid point in the parameter space. 
Ref.~\cite{Ajith:2007kx} proposed a family of Fourier domain templates of the form:
\be
u(f) \equiv {A}_{\rm eff}(f) \, e^{\romani\Psi_{\rm eff}(f)},
\label{eq:phenWave}
\ee
where the effective amplitude and phase are expressed as:
\ber
{A_{\rm eff}}(f) \equiv C
\left\{ \begin{array}{ll}
\left(f/\fmerg\right)^{-7/6}   & \textrm{if $f < \fmerg$}\\
\left(f/\fmerg\right)^{-2/3}   & \textrm{if $\fmerg \leq f < \fring$}\\
w \, {\cal L}(f,\fring,\sigma) & \textrm{if $\fring \leq f < \fcut$,}\\
\end{array} \right. \nonumber \\
\Psi_{\rm eff}(f) \equiv 2 \pi f t_0 + \varphi_0 + \frac{1}{\eta}\,\sum_{k=0}^{7} 
(x_k\,\eta^2 + y_k\,\eta + z_k) \,(\pi M f)^{(k-5)/3}\,.
\label{eq:phenWaveAmpAndPhase}
\eer
In the above expressions, $C$ is a numerical constant whose value depends on the 
sky-location and orientation of the binary as well as its physical parameters. 
For optimally-located and- oriented binaries, 
$C = \frac{M^{5/6}\,\fmerg^{-7/6}}{d\,\pi^{2/3}} \sqrt{\frac{5\,\eta}{24}}$. 
$t_0$ is the time of arrival of the signal at the detector, $\varphi_0$ the initial phase, 
${\cal L}(f,\fring,\sigma ) \equiv \left(\frac{1}{2 \pi}\right) \frac{\sigma}{(f-\fring)^2+\sigma^2/4}$
a Lorentzian function with width $\sigma$ centered around the frequency $\fring$, $w$ a 
normalisation constant chosen so as to make ${A}_{\rm eff}(f)$ continuous across the `transition' 
frequency $f_{\rm ring}$, and $f_{\rm merg}$ is the frequency at which the power-law changes from
$f^{-7/6}$ to $f^{-2/3}$. The phenomenological parameters 
$\fmerg, \fring, \sigma$ and $\fcut$ are written in terms of the total mass $M$ and symmetric 
mass ratio $\eta$ of the binary as
\ber
\pi M \fmerg =  a_0 \, \eta^2 + b_0 \, \eta + c_0  \,, \nonumber \\
\pi M \fring =  a_1 \, \eta^2 + b_1 \, \eta + c_1  \,, \nonumber \\
\pi M \sigma =  a_2 \, \eta^2 + b_2 \, \eta + c_2  \,, \nonumber \\
\pi M \fcut  =  a_3 \, \eta^2 + b_3 \, \eta + c_3. 
\label{eq:ampParams}
\eer
The coefficients $a_j, b_j, c_j,~j=0...3$ and $x_k,y_k,z_k,~k=0,2,3,4,6,7$ are unique for a given 
family of hybrid waveforms. The coefficients corresponding to the hybrid waveforms considered
here are tabulated in Table~\ref{tab:polCoeffs}. These are computed from
7 hybrid waveforms in the range $0.25 \geq \eta \geq 0.16$  produced
by matching the numerical waveforms (referred to as the `long NR waveforms' in Ref.~\cite{Ajith:2007kx}) 
produced by the Jena group with restricted 3.5PN TaylorT1 waveforms. 

\begin{table}[tbh]
    \begin{center}
        \begin{tabular}{cclclcll}
            \hline
            \hline
            Parameter &\vline& \multicolumn{1}{c}{$a_k$} &\vline& \multicolumn{1}{c}{$b_k$} &\vline& \multicolumn{1}{c}{$c_k$} \\
            \hline
                $\fmerg$  &\vline&   6.6389  $\times 10^{-1}$ &\vline&  -1.0321 $\times 10^{-1}$ &\vline& 1.0979 $\times 10^{-1}$ \\
    		$\fring$  &\vline&   1.3278                    &\vline&  -2.0642 $\times 10^{-1}$ &\vline& 2.1957 $\times 10^{-1}$ \\
    		$\sigma$  &\vline&   1.1383                 &\vline&  -1.7700 $\times 10^{-1}$ &\vline& 4.6834 $\times 10^{-2}$ \\
    		$\fcut$   &\vline&   1.7086                    &\vline&  -2.6592 $\times 10^{-1}$ &\vline& 2.8236 $\times 10^{-1}$ \\
             \hline
            Parameter &\vline& \multicolumn{1}{c}{$x_k$} &\vline& \multicolumn{1}{c}{$y_k$} &\vline& \multicolumn{1}{c}{$z_k$}  \\
            \hline
            $\psi_0$  &\vline&   -1.5829 $\times 10^{-1}$ &\vline&  8.7016 $\times 10^{-2}$  &\vline& -3.3382 $\times 10^{-2}$  \\
    	    $\psi_2$  &\vline&   3.2967  $\times 10^{1}$  &\vline& -1.9000 $\times 10^{1}$   &\vline&  2.1345  \\
    	    $\psi_3$  &\vline&   -3.0849 $\times 10^{2}$  &\vline&  1.8211 $\times 10^{2}$   &\vline& -2.1727 $\times 10^{1}$  \\
    	    $\psi_4$  &\vline&   1.1525  $\times 10^{3}$  &\vline& -7.1477 $\times 10^{2}$   &\vline&  9.9692 $\times 10^{1}$  \\
    	    $\psi_6$  &\vline&   1.2057  $\times 10^{3}$  &\vline& -8.4233 $\times 10^{2}$   &\vline&  1.8046 $\times 10^{2}$  \\
    	    $\psi_7$  &\vline&   0                        &\vline&  0                        &\vline& 0\\
	     \hline
            \hline
        \end{tabular}
        \caption{Coefficients describing the amplitude and phase of the phenomenological waveforms. See Eqs.
        (\ref{eq:phenWaveAmpAndPhase}) and (\ref{eq:ampParams}).}
        \label{tab:polCoeffs}
    \end{center}
\end{table}

\section{Assessing the efficiency of different searches}
\label{sec:searchCompare}

\begin{figure}[tbp]
\centering
\includegraphics[width=5.5in]{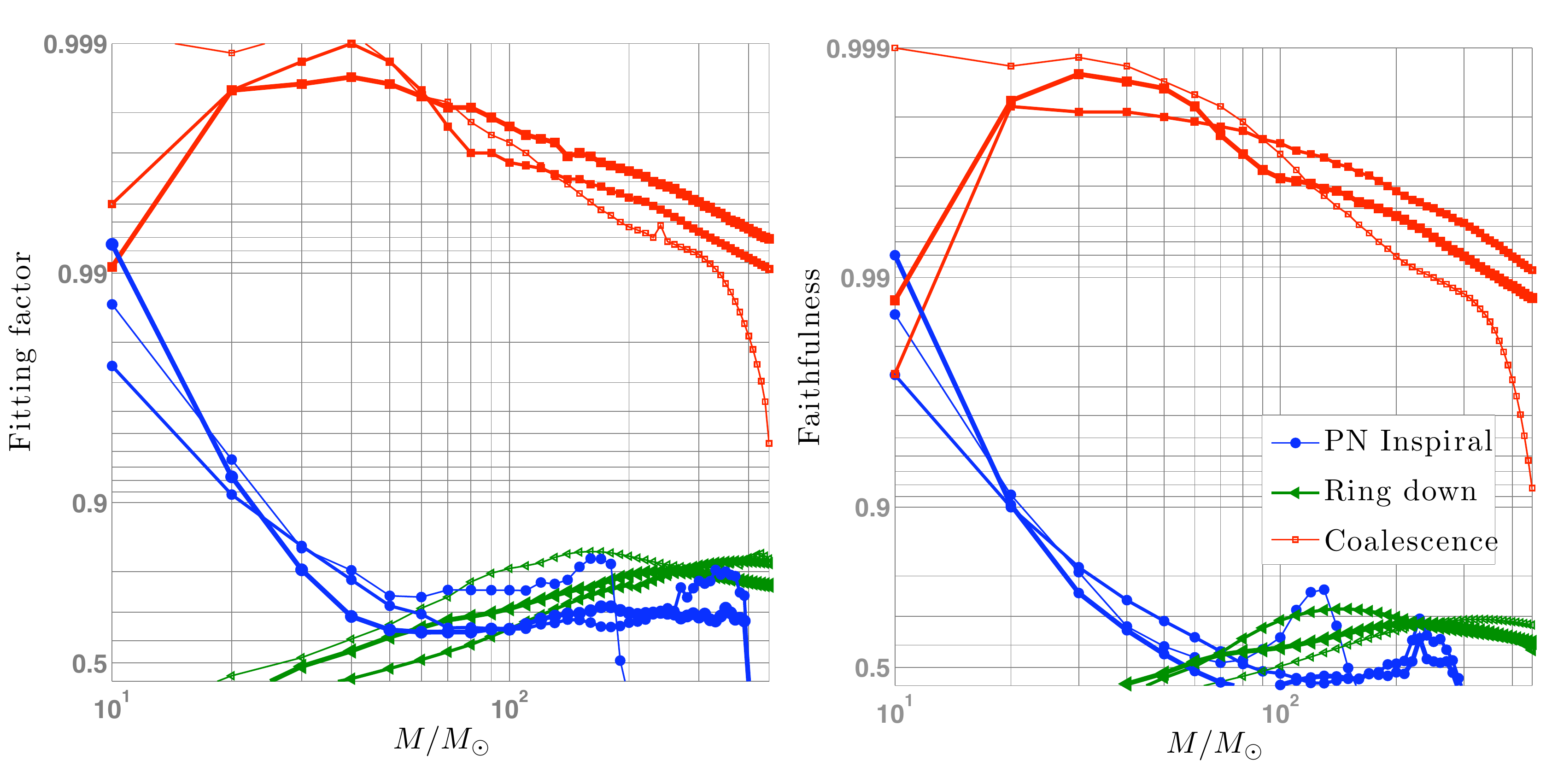}
\caption{Fitting factors (left plots) and faithfulness (right plots) of the PN inspiral, 
ring down and coalescence templates. Dots correspond to PN inspiral templates, triangles
to ring-down templates, and squares to coalescence templates. The overlaps are computed 
using three different noise spectra -- Initial LIGO (thin lines), Virgo (thicker lines) 
and Advanced LIGO (thickest lines). Horizontal axes report the total mass of the binary.}
\label{fig:FFandFaithfulness}
\end{figure}

The hybrid waveforms constructed in Section~\ref{sec:hybWave} can be used to estimate the efficiency
of different searches (including the one proposed in the previous section) in detecting signals from 
binary black-hole coalescences~\footnote{The obvious assumption involved is that the hybrid waveforms
are sufficiently close to the signals produced by nature. This is greatly dependent on the systematic 
errors in the hybrid waveforms.}. Ideally, this should be done by injecting a large number of hybrid 
waveforms from different binaries into the detector data and by estimating the fraction of the injections 
detected by each search. Different search groups have already started work in this direction~\cite{HybWaveInject}. 
This section presents a simple strategy to estimate the 
efficiency of different template families in detecting signals from binary black-hole coalescences. 
The template families being considered here are (i) restricted 3.5PN TaylorT1 inspiral templates 
truncated at the maximum binding-energy circular orbit~\cite{Blanchet:2002} (ii) black-hole ring-down 
templates proposed in Ref.~\cite{Creighton:1999pm}, and (iii) black-hole coalescence templates described 
in the previous section. 

We compute the \emph{fitting factors}~\cite{Apostolatos:1995pj} and \emph{faithfulness}~\cite{DIS98} 
of different template families with the hybrid waveforms. Fitting factor is the overlap of a template
waveform with the target signal maximised over both the \emph{intrinsic} ($M$ and $\eta$) and the 
\emph{extrinsic} ($t_0$ and $\varphi_0$) parameters of the template waveform, while faithfulness is 
the overlap maximised over only the \emph{extrinsic} parameters of the template. Faithfulness is a 
measure of how good the template waveform is in both detecting a signal and estimating its parameters. 
However, the fitting factor is aimed at finding whether or not a template family is good enough in detecting 
a signal without reference to its use in estimating the parameters.

The fitting factors and faithfulness of three different template families, using the hybrid waveforms 
as target signals, are plotted in Figure~\ref{fig:FFandFaithfulness}. Maximization over the intrinsic parameters
is performed with the aid of the Nelder-Mead downhill simplex algorithm. Low frequency cutoff is chosen 
to be equal to 40Hz for Initial LIGO and 20Hz for Virgo and Advanced LIGO. As expected, PN inspiral templates 
produce very good overlaps with the target signals in the low-mass regime (where inspiral is the dominant 
part), and ring-down templates produce good overlaps in the high-mass regime (where ring down is the 
dominant part). The black-hole coalescence templates continue to produce very good overlaps over the 
entire mass range. Note that, if we assume homogeneous and isotropic distribution of sources, the fraction 
of sources detectable by a template family is proportional to the cube of the fitting factor~\cite{Owen:1995tm}. 
The `event loss' due to mismatch between the template and the true signal is larger than the canonical 
10\% when the fitting factor is less than 0.965~\footnote{It should be noted that fitting factor is not the 
only consideration in an actual search strategy. Other factors such as computational cost and false alarm
rate also play a decisive role in the choice of a template family. However such a detailed study is out 
of the scope of this paper.}. 
This figure suggests that the template family proposed in Section~\ref{sec:phenTempl} can be used to search for 
binary black-hole coalescences over almost the entire mass range considered here losing no more 
than 10\% of the events that are detectable by optimal filtering. Binaries with total mass $\lesssim 15 M_\odot$ 
are detectable using PN inspiral templates with $< 10\%$ event loss, while the binaries in the mass range 
considered here cannot be detected with $< 10\%$ event loss using black-hole ring-down templates~\footnote{Note that the 
overlaps are maximised over the initial phase $\varphi_0$ of the ring down also, unlike what is proposed 
in Ref.~\cite{Creighton:1999pm}.}. Faithfulness of the coalescence templates is also almost always greater 
than 0.965 (comparable to the fitting factors), while that of the other templates are considerably smaller in 
general. This also means that the parameters estimated by the PN and ring-down templates will be biased 
significantly. This will be studied in detail in a forthcoming work~\cite{NRDApaper3}.

\begin{figure}[tbp]
\centering
\includegraphics[width=5.5in]{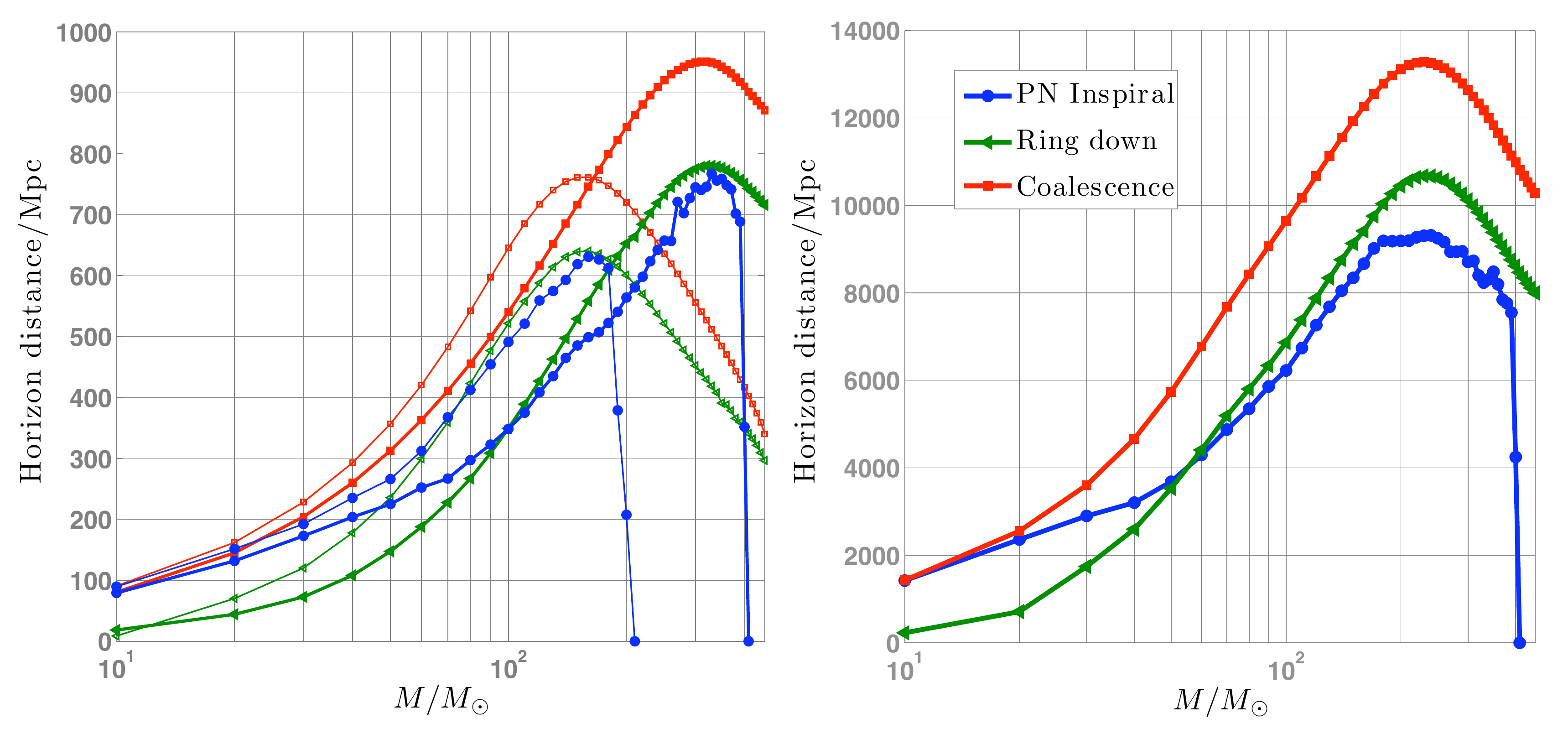}
\caption{Left plots shows the horizon distances of PN inspiral, ring down and coalescence
templates in the case of Initial LIGO (thin lines) and Virgo (thick lines) noise spectra. 
Right plots show the same for Advanced LIGO. Dots correspond to PN inspiral templates, 
triangles to ring-down templates, and squares to coalescence templates. Horizontal axes
report the total mass of the binary, and the vertical axes report the effective distance to 
optimally-oriented equal-mass binaries producing a \emph{sub-optimal} SNR of 8 at the 
detector output. The sharp drop in the PN horizon distance is a result of the (different) 
lower cutoff frequencies of the detectors.}
\label{fig:HorizonDistance}
\end{figure}

We can also calculate the `distance reach' of these searches. Since the template waveforms described 
in Eqs.(\ref{eq:phenWave}) and (\ref{eq:phenWaveAmpAndPhase}) are shown to be very close (fitting factors 
$>$ 0.95) to the hybrid waveforms, the effective distance $d_{\rm opt}$ to binaries producing a certain 
\emph{optimal} signal-to-noise ratio (SNR) $\rho$ at the detector can be computed analytically 
using these template waveforms, as:
\be
d_{\rm opt} = \frac{2}{\rho} \left[ \int_{f_{\rm low}}^{\fcut} \frac{A_{\rm eff}(f) \, \df}{S_h(f)}\right]^{1/2}.
\ee
Since the fitting factor (${\rm FF}$) is the fraction of the optimal SNR that can be achieved using a 
sub-optimal filter, the effective distance $d_{\rm subopt}$ to the (optimally-oriented) binaries producing 
a \emph{sub-optimal} SNR $\rho$ by a template family is given by $d_{\rm subopt} = d_{\rm opt} \, {\rm FF}$. 
Figure~\ref{fig:HorizonDistance} compares the effective distance to optimally-located and- oriented binaries 
producing a sub-optimal SNR of 8 at the detector output using the three different template families discussed 
above.  For the black-hole coalescence templates, the horizon 
distance reaches peak values of around 760 Mpc (at 150 $M_\odot$), 950 Mpc (325 $M_\odot$) and 13.3 Gpc 
(225 $M_\odot$) for Initial LIGO, Virgo and Advanced LIGO, respectively. For PN inspiral templates, the 
peak values are 630 Mpc (160 $M_\odot$), 770 Mpc (325 $M_\odot$) and 9.2 Gpc (250 $M_\odot$), while 
for the ring-down templates, the corresponding values are 640 Mpc (150 $M_\odot$), 780 Mpc (325 $M_\odot$) 
and 10.6 Gpc (225 $M_\odot$). It may be noted that, since the event rate is proportional to the cube of
the distance reach, a 20\% loss in the distance reach means a 50\% loss in the event rate. 

\section{Summary and future work}
\label{sec:Summary}

Recent progress in the theoretical modelling of coalescing binary black holes has important applications 
in the search for gravitational waves from binary black-hole coalescences. `Complete' gravitational 
waveforms can be constructed
by combining results from analytical and numerical calculations. These waveforms can be parametrised to 
produce analytical waveform templates which can be used to densely cover the parameter space of the binary
that will be searched over by matched-filtering techniques. This template family will allow us 
to coherently search over all the three stages of the binary. The advantages of this `coherent search' 
include improved SNR, and hence improved distance reach for the search, potential reduction of the 
false-alarm rate and improved parameter estimation.  
The application of the complete waveforms is not limited to template-based searches. For example, these 
waveforms can also be used in the burst searches, customised for detection of signals from coalescing 
black-hole binaries. In this case, a small bank of representative waveforms can be used to survey the 
parameter space of the binary~\cite{Klimenko:2007}.
The complete waveforms can also be used to estimate the efficiency of 
different search methods. A preliminary comparison of three different template-based searches is 
presented in this paper. More robust ways of comparing the efficiency of different searches is an 
ongoing effort~\cite{HybWaveInject}.

\section*{Acknowledgments}
I would like to thank the numerical relativity groups of Albert Einstein Institute and 
University of Jena for sharing their results of binary black-hole simulations, and all the members 
of the AEI-Jena NRDA Collaboration for very useful discussions. Conversations with Yanbei Chen, 
Sascha Husa, Mark Hannam, Badri Krishnan, Sergey Klimenko and Stas Babak were particularly valuable. 

\section*{References}
\bibliographystyle{iopart-num}
\bibliography{template}

\end{document}